# Urban Visual Appeal According to ChatGPT: Contrasting AI and Human Insights

**Milad Malekzadeh**
Digital Geography Lab, Department of Geosciences and Geography, University of Helsinki, Finland
milad.malekzadeh@helsinki.fi; https://orcid.org/0000-0003-2275-0497; https://twitter.com/MiladMzdh; https://www.linkedin.com/in/miladmalekzadeh/;

**Elias Willberg**
Digital Geography Lab, Department of Geosciences and Geography, University of Helsinki, Finland
elias.willberg@helsinki.fi; https://orcid.org/0000-0003-0159-0084; https://twitter.com/EliasW_; https://www.linkedin.com/in/elias-willberg-765a42122/;

**Jussi Torkko**
Digital Geography Lab, Department of Geosciences and Geography, University of Helsinki, Finland
jussi.torkko@helsinki.fi; https://orcid.org/0009-0004-2374-0981; https://www.linkedin.com/in/jtorkko/;

**Tuuli Toivonen**
Digital Geography Lab, Department of Geosciences and Geography, University of Helsinki, Finland
tuuli.toivonen@helsinki.fi; https://orcid.org/0000-0002-6625-4922; https://twitter.com/TuuliToivonen; https://www.linkedin.com/in/tuuli-toivonen-6b613913/;

## Highlights

- The study leverages GPT-4 model with 1800+ Street View images of Helsinki for comprehensive urban visual appeal analysis.
- AI and human ratings of urban visual appeal show strong alignment, with notable geographic variations.
- Residents' ratings show more spatial variation compared to non-residents due to personal experiences.
- Results suggest AI models need local context understanding to match human evaluative nuances.
- Findings emphasize hybrid AI-human approaches for effective urban planning and design decisions.

## Abstract

The visual appeal of urban environments significantly impacts residents' satisfaction with their living spaces and their overall mood, which in turn, affects their health and well-being. Given the resource-intensive nature of gathering evaluations on urban visual appeal through surveys or inquiries from residents, there is a constant quest for automated solutions to streamline this process and support spatial planning. In this study, we applied an off-the-shelf AI model to automate the analysis of urban visual appeal, using over 1,800 Google Street View images of Helsinki, Finland. By incorporating the GPT-4 model with specified criteria, we assessed these images through three criteria-based prompts. Simultaneously, 24 participants, categorised into residents and non-residents, were asked to rate the images. Our results demonstrated a strong alignment between GPT-4 and participant ratings, although geographic disparities were noted. Specifically, GPT-4 showed a preference for suburban areas with significant greenery, contrasting with participants who found these areas less appealing. Conversely, in the city centre and densely populated urban regions of Helsinki, GPT-4 assigned lower visual appeal scores than participant ratings. While there was general agreement between AI and human assessments across various locations, GPT-4 struggled to incorporate contextual nuances into its ratings, unlike participants, who considered both context and features of the urban environment. The study suggests that leveraging AI models like GPT-4 allows spatial planners to gather insights into the visual appeal of different areas efficiently, aiding decisions that enhance residents' and travellers' satisfaction and mental health. However, caution is necessary, particularly when interpreting results for suburban and densely-populated areas. While we used an off-the-shelf model, it is crucial to develop models specifically trained to understand the local context and provide insights into human perceptions of urban elements. Although AI models provide valuable insights, human perspectives are essential for a comprehensive understanding of urban visual appeal. This will ensure that planning and design decisions promote healthy living environments effectively.

**1. Introduction**

Urban environments have a profound role on our satisfaction, travel behaviour, and our health and well-being (1,2). Growing evidence shows how well-designed, pleasant urban environments that attract people to walk and cycle, can lead to a range of positive impacts, from higher physical activity and improved mood, to increased active travel, and economic vibrancy (2–5). Yet, it is also increasingly understood that unpleasant environments can inflict various negative emotions, including fear and anxiety, and be prone to adverse consequences such as crime, car dependence, and avoidance (1,6,7). Obviously, it is highly relevant for local planners and decision-makers to understand and locate such features when developing cities.

While the key components of attractive urban environments were identified long ago (8,9), and have been recognised by the research community (3,10–12), the operationalisation of these principles into meaningful and robust spatial indicators is highly dependent on the availability of data. In this respect, recent years have witnessed a dramatic increase in the availability of micro-scale data collected from the street level, representing the immediate urban environment. The emergence of street view imagery (SVI) in many cities has provided a rich source of data with which to assess the visual quality of streets (13). Together with rapidly developed computer vision techniques for object detection, SVIs have allowed researchers to capture detailed street features automatically, thereby overcoming some of the limitations of less detailed neighbourhood-level metrics or field-based audit data collection (14). A burgeoning literature has applied SVIs to a variety of urban use cases, including the assessment of walkability (15,16), pedestrian and cycling safety (17,18), pedestrian and cycling volume and behaviour (19,20), street greenery (21,22), microclimate (23,24), and physical disorder of streets (25,26).

From the planning perspective, the process of turning the visual information contained by SVIs into environmental indicators applicable to planning practice nevertheless remains a challenge. Existing literature on SVIs and environmental quality has largely focused on searching for the most important correlation between visual street features and travel behaviour, improving the accuracy of existing street quality indicators, and mapping the spatial distribution of distinct environmental features (e.g., visual complexity or street enclosure) in the study cities (13,27). Despite the obvious potential of SVIs, translating their potential into planning practice is a non-trivial issue. Common hardships include distilling multiple attributes into conceptually and methodologically robust but simple indicators, ensuring spatial coverage, high requirements for technical and methodological know-how, and the need for computational capacity. It is therefore necessary to continue the search for ways which will lower entry barriers and streamline the process of harnessing SVIs when evaluating urban environmental quality in planning.

Ensuring a high likelihood of finding these ways lies in the recent breakthroughs in artificial intelligence (AI). The emergence of Multimodal Large Language Models (MLLMs) like GPT-4 (Open AI) and Kosmos-2.5 (Microsoft) with capabilities to integrate the textual interaction capability with image analysis, holds great promise. The capacity of MLLMs to produce human-like text based on large amounts of data, opens new opportunities for applications requiring comprehensive analysis of both visual and textual information, such as the visual assessment of an urban area. Above all, the use of current MLLMs based on simple textual inputs can lower the barriers for using street view data in planning processes and can produce operable environmental quality indicators. However, it is still little known how well the results produced by MLLMs on the visual appeal reflect people's experience, which is a central requirement for their application. Overall, the potential of MLLMs in analysing the visual appeal of urban environments, remains largely underexplored.

For this study, we explored the potential of MLLMs to produce assessments of the visual appeal of urban environments. We applied an AI model to automate the analysis of over 1,800 Google Street View (GSV) images collected in Helsinki, Finland. By incorporating the GPT-4 model with urban environmental quality criteria from the literature, we assessed these images through the set of three input prompts, from simpler, to more complex. To compare the ratings of environmental quality that we obtained from the AI model, we asked 24 participants, categorised into residents and non-residents, to rate the visual appeal of the urban environments in the images. By comparing the ratings of the AI models and our participants statistically and spatially, we revealed the potential, as well as the limitations, of MLLMs, when applied to environmental quality assessment. Finally, we discuss the implications and limitations of our approach in planning.

## 2. AI in Sentiment and Multimodal Analysis

The advent of Large Language Models (LLMs) such as GPT-3 (28) and BERT (29) has not only revolutionised AI with sophisticated text generation and understanding, but also democratised interactions with AI technologies (30). These models enable users to execute commands, optimise and fine-tune AI responses, and engage in nuanced interactions without requiring deep AI expertise. This suite of LLMs illustrates the leap towards intuitive, accessible technology, transforming user interactions across various domains. This capability is particularly crucial for our study, allowing for the customisation of analysis criteria. Nonetheless, the application of LLMs is limited, because it lacks the ability to process visual media, essential for assessing urban visual appeal.

The emergence of MLLMs like GPT-4 (31) and Kosmos-2.5 (32,33), which integrate the textual interaction capability of LLMs with image analysis, presents a novel solution. Early studies in the field were dedicated to understanding and generating text based on multimodal inputs, focusing on how models interpret the relationship between visual elements and text. This research area benefited greatly from projects like BLIP-2 (Bootstrapping Language-Image Pre-training-2) (34), CLIP (Contrastive Language-Image Pretraining) (35), and LLaVA (Large Language-and-Vision Assistant) (36). As the field evolved, the scope of MLLMs broadened to include generating outputs specific to various modalities. These models emphasise the use of modality encoders, LLM backbones, and modality generators to process and generate multimodal content efficiently (37). Notable in this development is the flexibility in input representation, allowing for seamless integration of several data types into the LLM framework.

This advancement provides new options for applications requiring comprehensive analysis of both visual and textual information, such as an urban area's visual assessment, that can potentially benefit spatial planning. Some early studies have explored this field, including the study by Jongwiriyanurak et al. (38), which used LLaVA by prompting six questions to gather information on various factors considered critical in assessing motorcycle crash risks. Similarly, Liu et al. (39) employed CLIP to assess perceived walkability by analysing both tangible and subjective factors such as safety and attractiveness. Despite the progress made, the deployment of multimodal LLMs for a detailed analysis of urban visual appeal is still largely underexplored.

## 3. Determinants of Urban Visual Appeal

During recent decades, the determinants of urban visual appeal have become well established by an interdisciplinary research community. Within this literature, one of the key works is the book by Ewing et al., *Measuring Urban Design: Metrics for livable places* (40), which details a framework of metrics for measuring the quality of urban environments, as well as the definitions and measurement protocols to operationalise these metrics. This framework is grounded in the multidisciplinary understanding of how physical spaces and urban design qualities interact to influence individual reactions and behaviours, particularly walking behaviour, which is often a proxy for urban visual appeal.

The framework divides the relevant metrics into three main groups. The first group, *enduring physical features*, comprises elements such as sidewalk features for pedestrian activity, street design for traffic and activity, tree canopy and greenery, physical indicators of human activity, and permanent lighting. The second group encompasses *urban design qualities*, including imageability, legibility, human scale, transparency, linkage, complexity, and coherence. Thirdly, the last group extends the evaluation criteria to include *individual reactions*, reflecting personal and emotional responses to the urban environment.

In the literature, physical features like sidewalk width, street width, and tree canopy are directly observable and are believed to influence the more subjective urban design qualities (41–43). These features are often used in active transportation audit instruments, to measure the quality of the walking or bicycling environment (42–47).

On the other hand, urban design qualities, while influenced by these physical features, contribute to a cumulative effect on the experience of walking down a street that is greater than the sum of the parts. For instance, imageability, a concept popularised by Lynch (48), refers to the quality in a physical object which gives it a high probability of evoking a strong image in any given observer. It is what makes a space memorable and distinct. Similarly, qualities such as legibility, which is the ease with which a place can be recognised and organised into a coherent pattern, play a crucial role in how an individual perceives and engages with an urban space (48,49). The concept of transparency, derived from architecture and urban planning, refers to the literal and figurative visibility of a place. It affects how individuals perceive the openness and accessibility of a space, which is crucial for the sense of appealing (50–52). Complexity and coherence, on the other hand, reflect the visual richness and orderly arrangement of urban elements, which have been empirically linked to people's preference for and engagement with urban spaces (53). Moreover, enclosure, as described by Alexander (54) and Jacobs (52), refers to the creation of well-defined outdoor spaces with clear shapes and boundaries, akin to rooms, which evoke feelings of safety, definition, and memorability. Lastly, human scale and linkage refer to how the proportions of space and elements correspond with human dimensions, ensuring comfort (55), connectivity of different spaces (56), and facilitating movement and interaction (57).

Moreover, the inclusion of subjective reactions in the evaluation criteria acknowledges the multifaceted nature of urban design qualities. While physical features can be measured objectively, their influence on individual perceptions and behaviours is subjective, and can vary widely. As such, an evaluation should consider individual reactions, such as a sense of safety, comfort, and interest, which are personal yet pivotal components of an environment's pleasantness. As Talen stated, the field of urban analysis has yet to reach consensus on the most appropriate measures to use (58). The literature reveals a diversity of approaches, with various studies opting for singular measures, others for combinations, but without a universally accepted standard.

## 4. Methodology

In this study, we used street view imagery as the input for MLLMs from which to evaluate urban visual appeal, which was then compared against human participant ratings. First, to find the evaluation criteria to optimise the AI model for our study, we defined a set of criteria and determinants of urban visual appeal. Our criteria were based on Ewing et al. (40) (see section 4) to align with established theories and empirical evidence from the urban design literature. These criteria served as prompts in conjunction with visual data when engaging with the AI model. Initially, imagery data were collected and subjected to preliminary analysis with a subsample of the data to determine the most appropriate AI model. Following the preliminary analysis and model selection, participants were asked to rate the images, allowing for a comparative analysis between AI-generated and human assessments. Detailed procedural steps are provided in the following.

### 4.1. Study Area

Our study took place in Helsinki, the capital of Finland (Figure 1-a). Helsinki is a medium-sized city with a population of about 650,000 and a total land area of 214 km$^2$ (59) (Figure 1-b). The city comprises various types of urban fabric, including a densely built urban core at the tip of the peninsula with the highest population density in the county of around 5550 people per km$^2$ and the centre of economic, cultural, and social activity (Figure 1-c). Further away from the city centre are residential areas in the western, northern, and eastern parts of the city. The city, apart from its centre, is characterised by its greenery and multiple green spaces, which comprise approximately one-third of the total land area of Helsinki.

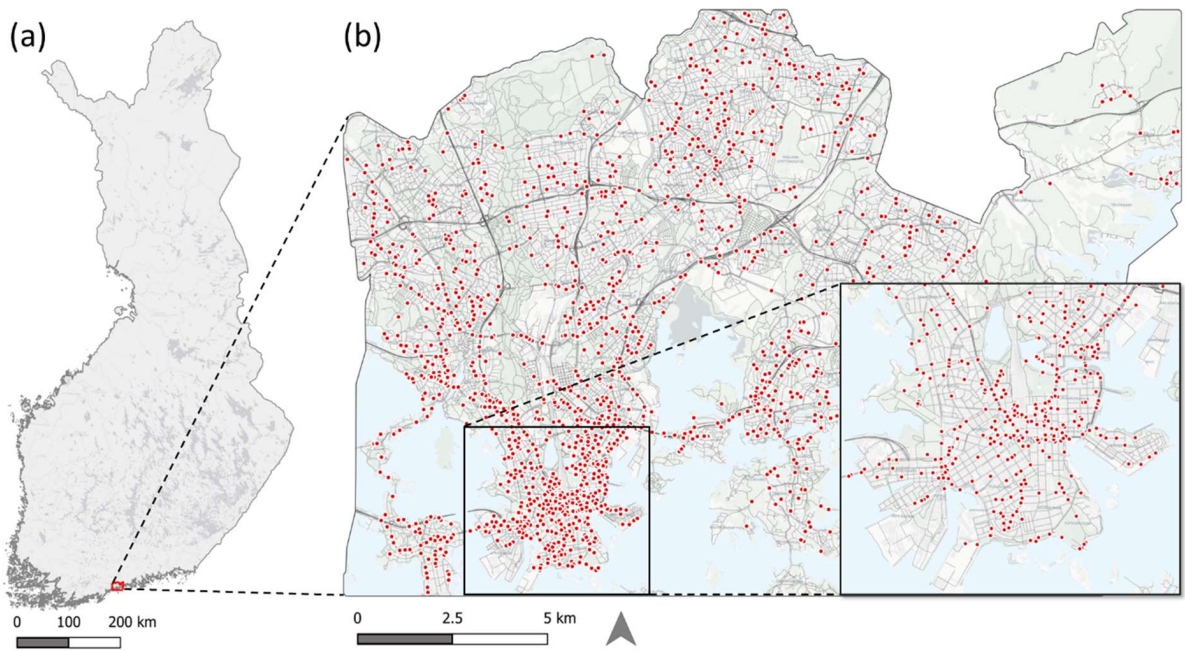

Figure 1 - (a) Location of Helsinki within Finland; (b) Helsinki's urban and suburban areas with red dots indicating the locations of Google Street View (GSV) images used in the study. Inset: A detailed view of Helsinki's urban core.

### 4.2. Acquisition of Urban Imagery

We acquired panoramic GSV images from Helsinki through Google Maps API with a key authorisation. The images had been collected between 2009 and 2017 over the study area (Figure 1-b). In the acquisition process, images were sampled at 20 metres intervals along the street network. Each GSV image included a time stamp referring to the month and year when the image was taken, as well as the coordinate location of the image. The size of the images was 640 × 640 with a field of view of 60° and pitch 0°. The panoramic 360° image that we used in the models and in the human evaluation, was composed of six directional images (0°, 60°, 120°, 180°, 240°, 300°) in each compass location, with 0° directed towards the north.

To select the images for AI and human evaluation, we first randomly sampled 1000 GSV image locations from the study area to ensure spatial coverage. To ensure that we did not miss important locations from the local residents' point of view, we consulted a survey by the City of Helsinki on the walkability of local neighbourhoods (60). We located images that were up to 50 metres to the important locations that the residents had mapped and added these to our sample. Finally, we removed duplicate images, which resulted into a total set 1967 images that we used as the input for the AI models and human evaluations.

### 4.3. AI Evaluation of Urban Imagery

In our preliminary analysis, we evaluated three MLLMs, including CLIP, BLIP, and GPT-4, to determine the most suitable model for our study. Results indicated that GPT-4 outperformed the others, yielding ratings closely aligned with those from a small group of participants (refer to Appendix A). Using OpenAI's API allowed us to automate the submission of prompts and to retrieve results for the extensive number of images, a necessity given the volume of data.

Drawing on the determinants of visual appeal outlined in section 4, we developed three distinct prompt types for GPT-4, ranging from simple to complex, to assess the influence of these criteria on the visual appeal scores. These prompts were iteratively refined during the preliminary analysis, to enhance the alignment between AI-generated assessments and participant ratings, thereby reducing discrepancies.

For the first prompt, we simply asked GPT-4 to rate the overall visual appeal on a scale from 1 (completely unappealing) to 7 (completely appealing), without specifying any criteria (Table 1). The second prompt incorporated a set of physical features. In the third prompt, we integrated urban design quality criteria, as well as subjective reactions, into the previous set of criteria. Each criterion in the second and third prompts included a brief definition. For all prompts, we instructed

GPT-4 to disregard temporary elements such as weather or passing vehicles, to ensure consistency in responses. We also developed two distinct prompts for querying ChatGPT, tailored to reflect either a local resident's or a non-resident's perspective, which could be typical or atypical in terms of local aesthetic and environmental viewpoints. Hereafter, we have referred to these prompts as Model-1, Model-2, and Model-3, respectively, with the suffixes 'LR' for local residents and 'NR' for non-residents. Each prompt was accompanied by a panoramic image of the area. The exact wording of these prompts is available in Appendix B.

| Prompt | Criteria |
|---|---|
| Model-1 (Prompt 1) | Overall visual appeal |
| Model-2 (Prompt 2) | **Enduring Physical Features** |
| |     - Sidewalk Features for Pedestrian Activity |
| |     - Street Design for Traffic and Activity |
| |     - Tree Canopy and Greenery |
| |     - Physical Indicators of Human Activity |
| |     - Permanent Lighting |
| Model-3 (Prompt 3) | **Enduring Physical Features** |
| |     - Sidewalk Features for Pedestrian Activity |
| |     - Street Design for Traffic and Activity |
| |     - Tree Canopy and Greenery |
| |     - Physical Indicators of Human Activity |
| |     - Permanent Lighting |
| | **Urban Design Qualities** |
| |     - Imageability |
| |     - Legibility |
| |     - Enclosure |
| |     - Human Scale |
| |     - Transparency |
| |     - Linkage |
| |     - Complexity |
| |     - Coherence |
| | **Subjective Reaction** |

Table 1 - Criteria used in each ChatGPT query prompt

For each criterion, GPT-4 was instructed to provide a single integer rating. In the case of the first prompt, only one overall rating was requested. In this study, we used a simple average without weighting the criteria, treating all as being equally important. However, future experiments could explore the application of weighted averages, potentially to achieve results that align more closely with participants' ratings.

### 4.4. Human Evaluation of Urban Imagery

To assess the ratings generated by GPT-4, we primarily recruited university students as participants, and personnel affiliated with the institutions of the authors. Recognising that familiarity with an area and associated memories can influence perceptions, we included both residents and non-residents in our participant pool, to evaluate these effects and to compare their ratings with those from GPT-4. The study involved 13 participants who were residents of Helsinki at the time of the experiment, and 11 non-residents. Participants received instructions via a concise guidance document. We intentionally did not highlight specific criteria, unlike the prompts used with GPT-4, allowing participants to rate the visual appeal based on their intuitive perception, as if they were physically present in the area. This approach encouraged a more subjective assessment rather than a detailed objective analysis of features. However, they were instructed to disregard any temporary features. For the exact wording of the instructions, please refer to Appendix C.

Given the substantial number of images involved and the time-intensive nature of the task, we requested that participants rate at least 500 images each, although they were permitted to rate more if they chose to. To ensure a balanced distribution of ratings and prevent any single image from being rated excessively or not at all, we strategically divided the batch of images. On average, each participant provided 1,014 ratings, accumulating a total of 24,349 ratings in total. Each image was rated at least 9 times. On average, each image received 11 ratings.

### 4.5. Adjusting Ratings for Comparative Analysis

Recognising the subjective nature of visual appeal, we noted significant variance in the average ratings provided by participants; the lowest individual average was 3.12, while the highest was 5.55. Such disparities indicated that comparing raw rating values could be problematic and not directly comparable. To address this, we adjusted ratings by subtracting the individual's average rating from each raw rating they provided. This approach highlighted the relative visual appeal of areas, showing whether they were perceived as better or worse than an individual's average ratings. To maintain consistency, we also adjusted the GPT-4 ratings by subtracting the mean rating of each prompt, thus adjusting the data across different evaluators and prompts.

To adjust for potential bias due of luminosity, we investigated whether luminosity influenced the ratings, based on the assumption that brighter and sunnier images might receive higher ratings. If this correlation had been significant, it would have been necessary to adjust for luminosity in our analysis. However, our findings showed no significant correlation between luminosity and the visual appeal ratings (Appendix D). Consequently, we did not adjust the images based on luminosity.

### 4.6. Statistical Analysis

We observed that the ratings from the first prompt resulted in a non-normal distribution (refer to Appendix E). Consequently, when comparing the distribution of ratings from the first prompt with those of residents and non-residents, we employed the Wilcoxon test. In contrast, since the distributions from prompts two and three were normally distributed, we applied T-tests for comparisons. Additionally, we calculated the Pearson correlation coefficient to assess the degree of correlation between the ratings from GPT-4 and those provided by residents or non-residents.

To evaluate the ratings spatially, we first assessed the overall spatial autocorrelation using Moran's I. We then calculated the differences between the ratings from GPT-4 and participants to further analyse these discrepancies using Moran's I. Subsequently, we identified clusters of these differences using local Moran's I. To pinpoint the hot spots (areas where the differences between GPT-4 and participant ratings are positive and significant) and cold spots (areas where these differences are negative and significant), we employed the Getis-Ord G* statistic.

## 5. Results

### 5.1. Descriptive findings

In the comparison between the local resident and non-resident groups of participants, we observed a similar pattern in the ratings (Table 2). Local residents showed a slightly lower standard deviation, but a broader range of values compared to non-residents. For the GPT-4 Model-1 (LR and NR), the standard deviation was higher, and the range of values broader than those observed in participant ratings, with both models displaying a tendency towards lower median values. This suggests that despite a general trend towards lower ratings, the values above the average were considerably higher for GPT-4 compared to the participants. When examining GPT-4 Models 2 and 3 (LR and NR), the standard deviations aligned closely with those of participants, but the range of values was broader, especially on the lower, negative values. Nevertheless, the values for the second and third quartiles remained similar across all models and participant ratings.

| Ratings         | Mean | std  | Min   | 25%   | 50%   | 75%  | Max  |
|-----------------|------|------|-------|-------|-------|------|------|
| Local Residents | 0    | 0.72 | -2.37 | -0.49 | 0.01  | 0.52 | 1.92 |
| Non-Residents   | 0    | 0.75 | -2.17 | -0.51 | 0.07  | 0.56 | 1.96 |
| Model-1 LR      | 0    | 1.00 | -2.67 | -0.83 | -0.83 | 1.00 | 2.84 |
| Model-2 LR      | 0    | 0.76 | -3.97 | -0.28 | 0.17  | 0.51 | 1.99 |
| Model-3 LR      | 0    | 0.76 | -3.76 | -0.37 | 0.15  | 0.52 | 1.78 |
| Model-1 NR      | 0    | 1.00 | -4.50 | -0.75 | -0.75 | 1.11 | 1.11 |
| Model-2 NR      | 0    | 0.76 | -3.84 | -0.30 | 0.16  | 0.48 | 1.82 |
| Model-3 NR | 0 | 0.75 | -3.70 | -0.36 | 0.15 | 0.52 | 1.80 |

Table 2 - Summary statistics for GPT-4 models and participant ratings

In analysing the distribution of ratings between GPT-4 models and participants using Wilcoxon and T-tests, we found no significant differences, as indicated by the p-values (Table 3). This similarity in distributions suggests that the ratings from the GPT-4 models align closely with those provided by participants, with no statistically significant variations between the groups.

| Ratings | Statistic | p-value |
|---|---|---|
| | **Local Residents** | |
| Model-1 LR (Wilcoxon) | 805480.0 | 0.50 |
| Model-2 LR (T-Test) | 0.95 | 0.33 |
| Model-3 LR (T-Test) | 0.96 | 0.33 |
| | **Non-Residents** | |
| Model-1 NR (Wilcoxon) | 808881.0 | 0.60 |
| Model-2 NR (T-Test) | 0.89 | 0.36 |
| Model-3 NR (T-Test) | 0.92 | 0.35 |

Table 3 - Statistical comparison of GPT-4 models and participant ratings' distributions (significance level = 0.05)

Analysis of Pearson's R correlation values between the ratings from GPT-4 models and participants reveals a moderate positive correlation, with the highest value being 0.54. The highest correlation within the local residents' group can be observed with GPT-4 Model-1 (Figure 2-a), while for the non-residents group, GPT-4 Model-3 shows the highest correlation (Figure 2-b). GPT-4 Model-2 and Model-3 were highly correlated with each other, whereas this high correlation did not hold between these two models and GPT-4 Model-1. However, these differences are slight and not substantial, all remaining below 0.06. This indicates minimal variation among the GPT-4 models in terms of their correlation levels with participant ratings.

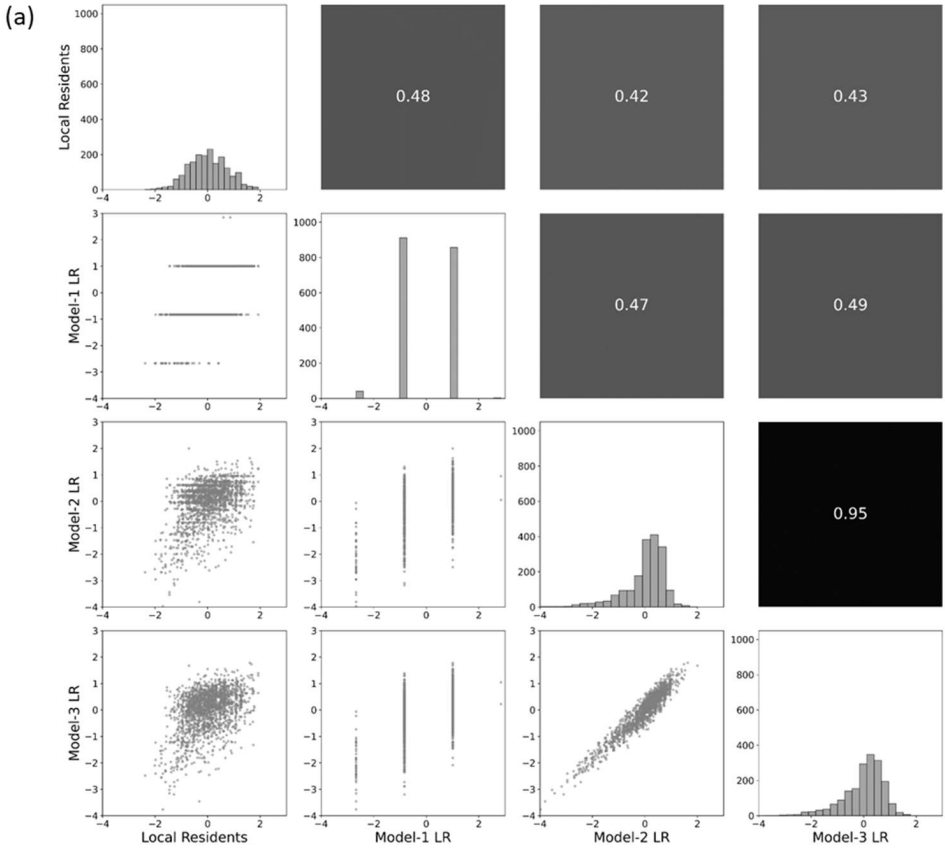

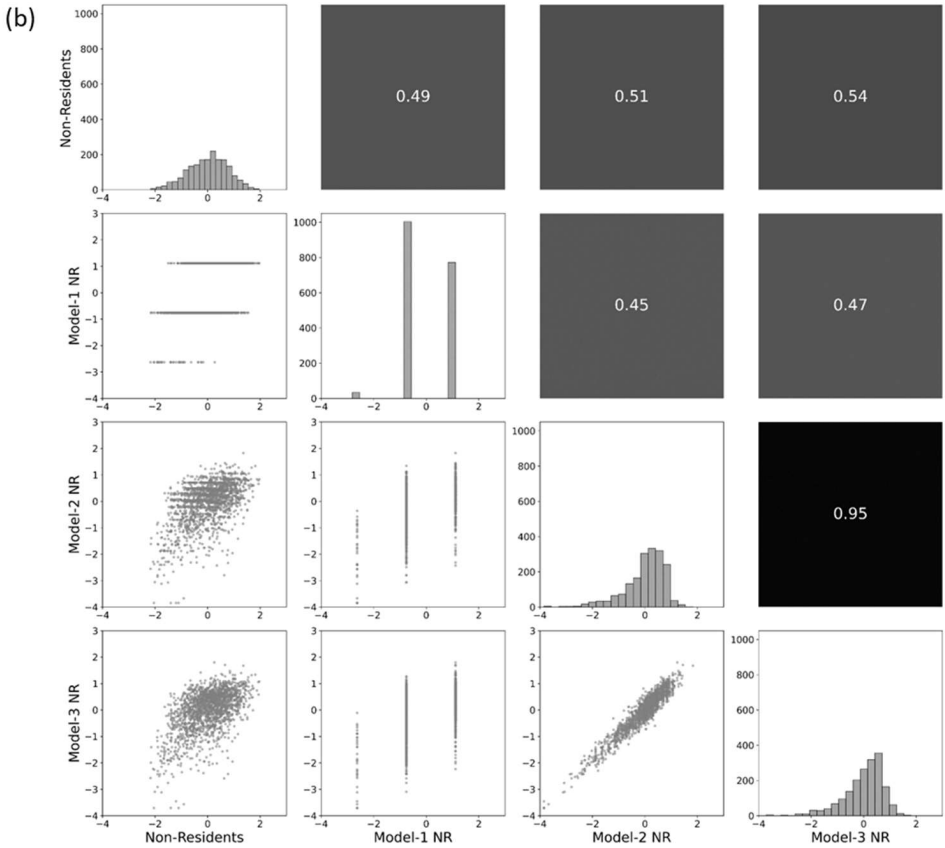

Figure 2 - Top half: Correlation heatmaps of Pearson's R values. Bottom half: Scatter plots. Diagonal: Histograms. The figure shows ratings derived from (a) GPT-4 LR models and local resident participants, and (b) GPT-4 NR models and non-resident participants.

## 5.2. Spatial Statistics

To explore spatial autocorrelation within our dataset, we first analysed the global Moran's I for each set of ratings (Table 4). The results indicated a positive spatial autocorrelation across all observations. Ratings from non-resident participants showed the highest level of spatial autocorrelation. This suggests that participants have more consistent and generalised perceptions of an urban area when it is unfamiliar to them. In contrast, residents, who have detailed and varied understanding based on their personal experiences and familiarity with the area, are likely to use more diverse criteria in their ratings, leading to lower spatial autocorrelation.

| Ratings | Moran's I | P-Value |
| --- | --- | --- |
| Local Residents | 0.26 | 0.001 |
| Non-Residents | 0.39 | 0.001 |
| Model-1 LR | 0.19 | 0.001 |
| Model-2 LR | 0.26 | 0.001 |
| Model-3 LR | 0.29 | 0.001 |
| Model-1 NR | 0.16 | 0.001 |
| Model-2 NR | 0.23 | 0.001 |
| Model-3 NR | 0.26 | 0.001 |

Table 4 - Spatial Autocorrelation (Moran's I) of ratings by local residents, non-residents, and GPT-4 models (significance level = 0.05).

Comparing the spatial autocorrelation of ratings derived from different GPT-4 models, we observe that the more complex the model (i.e., the more criteria in the prompt), the higher the spatial autocorrelation (Table 4). This result can be interpreted in two contradictory ways. First, simple models may apply fewer criteria and lack the sophistication to analyse complex urban features consistently, leading to more varied ratings, while more complex models apply more criteria uniformly, resulting in more homogenised evaluations and increased spatial autocorrelation. Alternatively, the additional criteria in more complex models might neutralise each other, leading to moderate and more similar ratings. This is supported by the descriptive analysis in Table 2, in which the range of the middle 50 percent of the data in Model-1 for both local residents and non-residents was higher than in Model-2 and Model-3. However, this analysis alone is insufficient to determine which interpretation is correct. To evaluate these views further, we also examined the spatial autocorrelation of differences between pairs of ratings (model-derived versus participant-derived) using Moran's I to evaluate the compatibility of these ratings with those of participants.

This analysis revealed that local residents' ratings exhibit slightly lower spatial autocorrelation compared to non-residents' ratings, indicating greater spatial variation in differences between local residents' ratings and those from GPT-4 models (Table 5). Comparing the GPT-4 models, we could observe a significant increase in spatial autocorrelation from Model-1 to Model-2 and Model-3, with the latter two models showing almost three times more spatial autocorrelation. This increase could be attributed to the specific criteria in Model-2 and Model-3, causing these models to treat areas that share partly similar characteristics uniformly.

To evaluate this further, we used Getis-Ord G* (Figure 3) and Local Moran's I (Appendix F) analyses. The results showed a clear pattern of more hot spots in suburban and rural areas in which GPT-4 model ratings were higher than participant ratings, while the reverse pattern was observed in densely populated urban areas. Additionally, we found that simpler models and non-residents had fewer hot/cold spots, compared to more complex models and local residents, indicating lower local spatial autocorrelation in differences between simpler models and non-residents' ratings.

| Ratings | Moran's I | p-value |
|---|---|---|
| **Local Residents** | | |
| Model-1 LR | 0.06 | 0.001 |
| Model-2 LR | 0.19 | 0.001 |
| Model-3 LR | 0.20 | 0.001 |
| **Non-Residents** | | |
| Model-1 NR | 0.08 | 0.001 |
| Model-2 NR | 0.21 | 0.001 |
| Model-3 NR | 0.23 | 0.001 |

Table 5 - Spatial Autocorrelation (Moran's I) of differences between GPT-4 model ratings and participant ratings (significance level = 0.05).

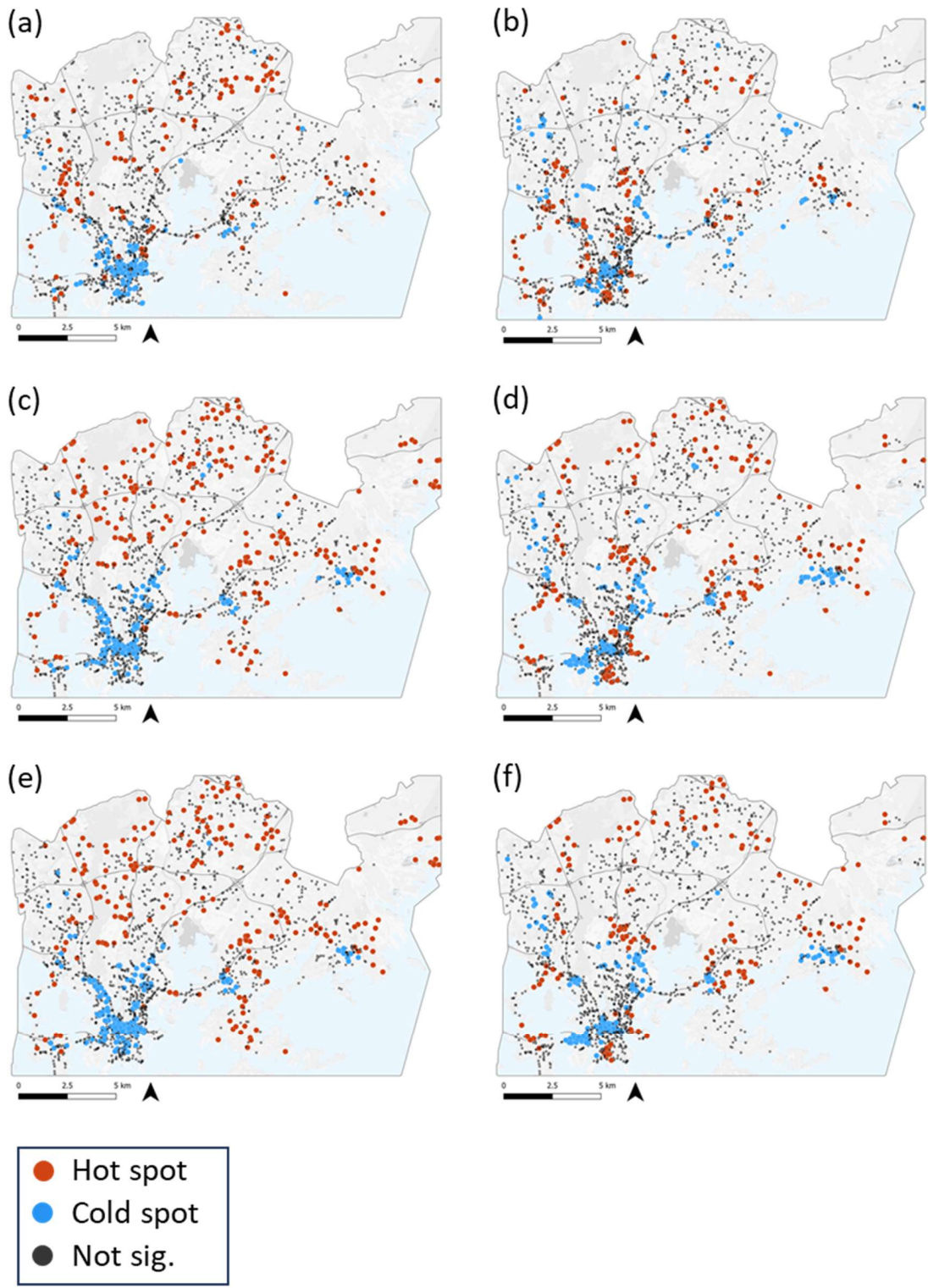

Figure 3 - Getis-Ord Gi hot spot analysis of the differences between GPT-4 model ratings and participant ratings (local residents and non-residents) for various models. Panels (a), (c), and (e)

show the results for Model-1, Model-2, and Model-3 respectively, for local residents, while panels (b), (d), and (f) correspond to the same models for non-residents. Red dots represent significant hot spots, where GPT-4 model ratings are higher than participant ratings and are clustered. Blue dots represent significant cold spots where participant ratings are higher than GPT-4 model ratings and are clustered. Black dots indicate areas with no significant clustering.

## 6. Discussion

Attractive urban environments are associated with a range of positive impacts, including citizen satisfaction, well-being, and sustainable travel behaviours (1,2). Advances in artificial intelligence models, combined with visual data covering urban areas, have made it possible to analyse the urban attractiveness in unprecedented ways. However, the gap between advances in mapping and planning practice remains wide, due to the complexity and resource-intensity of the novel methods. In this study, we explored the applicability of an off-the-shelf AI model with simple text commands in producing reliable spatial estimates of urban visual appeal from Street View Imagery.

Our findings revealed a general alignment between the AI-generated ratings and human evaluations on urban appeal, but also notable contextual differences. GPT models generally assigned higher ratings to suburban areas, and lower ratings to densely populated urban areas, compared to the human participants. This discrepancy may arise from the models' inability to grasp the contextual and cultural nuances that humans use when evaluating urban environments. Humans might find densely-populated areas appealing due to their vibrant social and economic activities, which are integral to their daily lives and enhance the urban quality of life (61), despite these areas being less green or visually complex. In Helsinki, GPT models underestimated the visual appeal, especially in the urban core with its dense population, active cultural and economic activity, and historical architecture, but less greenery. Conversely, suburban and rural areas, while often perceived as more visually appealing by the models due to their greenery and open spaces, might lack the dynamic elements that contribute to human satisfaction in urban settings, as the results indicated in Helsinki.

The analysis indicated that local residents exhibited lower spatial autocorrelation in their ratings compared to non-residents. This finding suggests that residents' evaluations are more diverse and influenced by their personal experiences and attachments to specific areas. In contrast, non-residents, who lack such personal connections, tend to rate urban areas based on more uniform and generalised criteria. This difference highlights how familiarity and a sense of place can shift evaluative criteria and perceptions (62). When living in an area, the daily interactions and memories associated with specific locations play a significant role in shaping one's evaluation of those places.

The analysis of spatial autocorrelation, both globally and locally, along with the differences between GPT model ratings and participant ratings, indicates that simpler models (Model-1) exhibited lower spatial autocorrelation compared to more complex models (Model-2 and Model-3). This suggests that simpler models, which lack specific evaluative criteria in their prompts, tend to provide more random and less consistent ratings. It should be noted that the complexity of human perception, which encompasses emotional, cultural, and experiential factors, presents a significant challenge for AI models (63). While current MLLMs like GPT-4 offer a promising start, they require guided prompting with certain criteria and/or fine-tuning to approach the depth of human evaluative processes.

An important reason for choosing off-the-shelf models was to democratise the use of AI for researchers and users who are not AI experts. These models eliminate the need for users to train the model, which often requires large amounts of training data and computational resources.

Instead, these models offer a straightforward and accessible way to utilise AI without extensive technical expertise. While we initially considered using other models for the analysis, ChatGPT outperformed the alternatives. ChatGPT is widely accessible and familiar to many users (64), making our workflow more user-friendly and reproducible. By providing our prompts, we have enabled others to modify and experiment with the criteria easily, engaging with ChatGPT in a conversational manner. This accessibility is crucial for ensuring that advanced AI tools can be used broadly in urban planning and design without requiring extensive technical expertise.

Despite its advantages, the use of ChatGPT comes with limitations. It is proprietary model and has usage constraints, even in its premium version (31). This can be a limiting factor for users needing to process numerous requests in a short period, as demonstrated in our workflow. Consequently, cities with extensive street view imagery, like Helsinki, would require a sampling algorithm to optimise the number of images for assessment, minimising the cost and time of API usage. Future research should focus on developing open-access models tailored to urban landscape analysis, allowing users to fine-tune and adapt these models without associated costs.

An important limitation of this research is that the results cannot be fully compared to the actual experiences of people physically present in an area. While this workflow and method can highlight areas with high or low visual appeal, the ratings are based on images rather than the real-life experience of being in a space. Capturing a multisensory, real-life experience in a relatively low resolution 360-degree image is impossible. Real-life experiences involve a combination of visual, auditory, tactile, and even olfactory stimuli that images alone cannot convey (65–67). Consequently, the ratings generated from these images only correspond to a partial representation of the actual environment.

## 7. Conclusion

While AI models like GPT-4 offer significant potential for streamlining the evaluation of urban visual appeal, our study highlights the need to incorporate human perspectives to capture the full range of contextual and experiential nuances. AI models can serve as an increasingly valuable tool for preliminary assessments, identifying areas that may require further human investigation. This approach can reduce operational costs and time expended by urban planners and designers, providing a more efficient pathway to understanding urban environments. However, caution should be exercised when relying solely on AI models for policymaking decisions, especially in areas in which human experiences and perceptions play a crucial role. The results of AI assessments should complement, rather than replace, human evaluations. Future research should focus on enhancing AI models to mimic human perception better, by integrating more sophisticated and nuanced criteria.

Our work demonstrates the potential for AI to aid in the assessment of urban landscapes, but it also underscores the limitations of current technology. The ongoing development of AI models that can better understand and replicate human experiences will be critical for their effective application in urban planning. Existing survey materials could be leveraged to build a training pipeline, enhancing AI's ability to provide meaningful insights tailored to the local context. Ultimately, a hybrid approach that leverages both AI and human insights will provide the most comprehensive understanding of urban visual appeal, ensuring that planning and design decisions effectively promote healthy and satisfying living environments.

**Disclosure statement**

No potential conflict of interest was reported by the authors.

**Ethics approval statement**

This study did not require ethics approval.

**Funding statement**

This study is part of the GREENTRAVEL project (2023-2027) funded by the European Union (ERC, project 101044906). Views and opinions expressed are however those of the authors only and do not necessarily reflect those of the European Union or the European Research Council Executive Agency. Neither the European Union nor the granting authority can be held responsible for them.

**References**

1.  Giles-Corti B, Vernez-Moudon A, Reis R, Turrell G, Dannenberg AL, Badland H, et al. City planning and population health: a global challenge. The lancet. 2016;388(10062):2912–24.

2.  Nieuwenhuijsen MJ. Urban and transport planning pathways to carbon neutral, liveable and healthy cities; A review of the current evidence. Environ Int [Internet]. 2020;140:105661. Available from: https://www.sciencedirect.com/science/article/pii/S0160412020302038

3.  Saelens BE, Handy SL. Built environment correlates of walking: a review. Med Sci Sports Exerc. 2008;40(7 Suppl):S550.

4.  St-Louis E, Manaugh K, van Lierop D, El-Geneidy A. The happy commuter: A comparison of commuter satisfaction across modes. Transp Res Part F Traffic Psychol Behav [Internet]. 2014;26:160–70. Available from: https://www.sciencedirect.com/science/article/pii/S1369847814001107

5.  van Wee B, Ettema D. Travel behaviour and health: A conceptual model and research agenda. J Transp Health [Internet]. 2016;3(3):240–8. Available from: https://www.sciencedirect.com/science/article/pii/S2214140516302006

6.  Carver A, Timperio A, Crawford D. Playing it safe: The influence of neighbourhood safety on children's physical activity—A review. Health Place [Internet]. 2008;14(2):217–27. Available from: https://www.sciencedirect.com/science/article/pii/S1353829207000536

7.  de Jong T, Fyhri A. Spatial characteristics of unpleasant cycling experiences. J Transp Geogr [Internet]. 2023;112:103646. Available from: https://www.sciencedirect.com/science/article/pii/S0966692323001187

8.  Gehl J. Life between buildings: Using public space. Island Press; 1987.

9.  Jacobs J. The death and life of great American cities. 1961. New York: Vintage. 1992;321:9783839413272–099.

10. Giles-Corti B, Donovan RJ. Relative influences of individual, social environmental, and physical environmental correlates of walking. Am J Public Health. 2003;93(9):1583–9.

11. Ewing R, Handy S. Measuring the Unmeasurable: Urban Design Qualities Related to Walkability. J Urban Des (Abingdon) [Internet]. 2009 Feb 1;14(1):65–84. Available from: https://doi.org/10.1080/13574800802451155

12. Forsyth A. What is a walkable place? The walkability debate in urban design. URBAN DESIGN International [Internet]. 2015;20(4):274–92. Available from: https://doi.org/10.1057/udi.2015.22

13. Biljecki F, Ito K. Street view imagery in urban analytics and GIS: A review. Landsc Urban Plan. 2021;215:104217.

14. Ki D, Chen Z, Lee S, Lieu S. A novel walkability index using google street view and deep learning. Sustain Cities Soc [Internet]. 2023;99:104896. Available from: https://www.sciencedirect.com/science/article/pii/S2210670723005073

15. Li Y, Yabuki N, Fukuda T. Measuring visual walkability perception using panoramic street view images, virtual reality, and deep learning. Sustain Cities Soc [Internet]. 2022;86:104140. Available from: https://www.sciencedirect.com/science/article/pii/S221067072200453X

16. Nagata S, Nakaya T, Hanibuchi T, Amagasa S, Kikuchi H, Inoue S. Objective scoring of streetscape walkability related to leisure walking: Statistical modeling approach with semantic segmentation of Google Street View images. Health Place [Internet]. 2020;66:102428. Available from: https://www.sciencedirect.com/science/article/pii/S1353829220302720

17. Mooney SJ, Wheeler-Martin K, Fiedler LM, LaBelle CM, Lampe T, Ratanatharathorn A, et al. Development and validation of a Google Street View pedestrian safety audit tool. Epidemiology. 2020;31(2):301–9.

18. Hamim OF, Ukkusuri S V. Towards safer streets: A framework for unveiling pedestrians' perceived road safety using street view imagery. Accid Anal Prev [Internet]. 2024;195:107400. Available from: https://www.sciencedirect.com/science/article/pii/S0001457523004475

19. Chen L, Lu Y, Sheng Q, Ye Y, Wang R, Liu Y. Estimating pedestrian volume using Street View images: A large-scale validation test. Comput Environ Urban Syst [Internet]. 2020;81:101481. Available from: https://www.sciencedirect.com/science/article/pii/S0198971519304351

20. Liu J, Ettema D, Helbich M. Street view environments are associated with the walking duration of pedestrians: The case of Amsterdam, the Netherlands. Landsc Urban Plan [Internet]. 2023;235:104752. Available from: https://www.sciencedirect.com/science/article/pii/S0169204623000713

21. Li X, Zhang C, Li W, Ricard R, Meng Q, Zhang W. Assessing street-level urban greenery using Google Street View and a modified green view index. Urban For Urban Green [Internet]. 2015;14(3):675–85. Available from: https://www.sciencedirect.com/science/article/pii/S1618866715000874

22. Ye Y, Richards D, Lu Y, Song X, Zhuang Y, Zeng W, et al. Measuring daily accessed street greenery: A human-scale approach for informing better urban planning practices. Landsc Urban Plan [Internet]. 2019;191:103434. Available from: https://www.sciencedirect.com/science/article/pii/S0169204618309940

23. Gong FY, Zeng ZC, Ng E, Norford LK. Spatiotemporal patterns of street-level solar radiation estimated using Google Street View in a high-density urban environment. Build Environ [Internet]. 2019;148:547–66. Available from: https://www.sciencedirect.com/science/article/pii/S0360132318306437

24. Sun Q (Chayn), Macleod T, Both A, Hurley J, Butt A, Amati M. A human-centred assessment framework to prioritise heat mitigation efforts for active travel at city scale. Science of The Total Environment [Internet]. 2021;763:143033. Available from: https://www.sciencedirect.com/science/article/pii/S0048969720365633

25. Mayne SL, Jose A, Mo A, Vo L, Rachapalli S, Ali H, et al. Neighborhood disorder and obesity-related outcomes among women in Chicago. Int J Environ Res Public Health. 2018;15(7):1395.

26. Keralis JM, Javanmardi M, Khanna S, Dwivedi P, Huang D, Tasdizen T, et al. Health and the built environment in United States cities: measuring associations using Google Street View-derived indicators of the built environment. BMC Public Health [Internet]. 2020;20(1):215. Available from: https://doi.org/10.1186/s12889-020-8300-1

27. Liu L, Sevtsuk A. Clarity or confusion: A review of computer vision street attributes in urban studies and planning. Cities [Internet]. 2024;150:105022. Available from: https://www.sciencedirect.com/science/article/pii/S0264275124002361

28. Brown T, Mann B, Ryder N, Subbiah M, Kaplan JD, Dhariwal P, et al. Language models are few-shot learners. Adv Neural Inf Process Syst. 2020;33:1877–901.

29. Devlin J, Chang MW, Lee K, Toutanova K. Bert: Pre-training of deep bidirectional transformers for language understanding. arXiv preprint arXiv:181004805. 2018;

30. Bilgram V, Laarmann F. Accelerating innovation with generative AI: AI-augmented digital prototyping and innovation methods. IEEE Engineering Management Review. 2023;

31. Achiam J, Adler S, Agarwal S, Ahmad L, Akkaya I, Aleman FL, et al. Gpt-4 technical report. arXiv preprint arXiv:230308774. 2023;

32. Huang S, Dong L, Wang W, Hao Y, Singhal S, Ma S, et al. Language is not all you need: Aligning perception with language models. Adv Neural Inf Process Syst. 2024;36.

33. Peng Z, Wang W, Dong L, Hao Y, Huang S, Ma S, et al. Kosmos-2: Grounding multimodal large language models to the world. arXiv preprint arXiv:230614824. 2023;

34. Li J, Li D, Savarese S, Hoi S. Blip-2: Bootstrapping language-image pre-training with frozen image encoders and large language models. In: International conference on machine learning. PMLR; 2023. p. 19730–42.

35. Radford A, Kim JW, Hallacy C, Ramesh A, Goh G, Agarwal S, et al. Learning transferable visual models from natural language supervision. In: International conference on machine learning. PMLR; 2021. p. 8748–63.

36. Liu H, Li C, Wu Q, Lee YJ. Visual instruction tuning. Adv Neural Inf Process Syst. 2024;36.

37. Zhang D, Yu Y, Li C, Dong J, Su D, Chu C, et al. Mm-llms: Recent advances in multimodal large language models. arXiv preprint arXiv:240113601. 2024;

38. Jongwiriyanurak N, Zeng Z, Wang M, Haworth J, Tanaksaranond G, Boehm J. Framework for Motorcycle Risk Assessment Using Onboard Panoramic Camera (Short Paper). In: 12th International Conference on Geographic Information Science (GIScience 2023). Schloss Dagstuhl-Leibniz-Zentrum für Informatik; 2023.

39. Liu X, Haworth J, Wang M. A New Approach to Assessing Perceived Walkability: Combining Street View Imagery with Multimodal Contrastive Learning Model. In: Proceedings of the 2nd ACM SIGSPATIAL International Workshop on Spatial Big Data and AI for Industrial Applications. 2023. p. 16–21.

40. Ewing R, Clemente O, Neckerman KM, Purciel-Hill M, Quinn JW, Rundle A. Measuring urban design: Metrics for livable places. Vol. 200. Springer; 2013.

41. Clifton KJ, Livi Smith AD, Rodriguez D. The development and testing of an audit for the pedestrian environment. Landsc Urban Plan [Internet]. 2007;80(1):95–110. Available from: https://www.sciencedirect.com/science/article/pii/S0169204606001101

42. Pikora TJ, Bull FCL, Jamrozik K, Knuiman M, Giles-Corti B, Donovan RJ. Developing a reliable audit instrument to measure the physical environment for physical activity. Am J Prev Med [Internet]. 2002;23(3):187–94. Available from: https://www.sciencedirect.com/science/article/pii/S0749379702004981

43. Wimbardana R, Tarigan AK, Sagala S. Does a pedestrian environment promote walkability? Auditing a pedestrian environment using the pedestrian environmental data scan instrument. J Reg City Plan. 2018;29:57–66.

44. Day K, Boarnet M, Alfonzo M, Forsyth A. The Irvine–Minnesota inventory to measure built environments: development. Am J Prev Med. 2006;30(2):144–52.

45. Emery J, Crump C, Bors P. Reliability and validity of two instruments designed to assess the walking and bicycling suitability of sidewalks and roads. American Journal of Health Promotion. 2003;18(1):38–46.

46. Khisty CJ. Evaluation of pedestrian facilities: beyond the level-of-service concept. Transp Res Rec. 1994;45.

47. Landis BW. Bicycle interaction hazard score: a theoretical model. Transp Res Rec. 1994;1438:3–8.

48. Lynch K. The image of the environment. The image of the city. 1960;11:1–13.

49. Evans GW, Smith C, Pezdek K. Cognitive maps and urban form. Journal of the American Planning Association. 1982;48(2):232–44.

50. Ewing R, Handy S. Measuring the unmeasurable: Urban design qualities related to walkability. J Urban Des (Abingdon). 2009;14(1):65–84.

51. Arnold HF. Trees in urban design. Trees in urban design. 1980;

52. Jacobs AB. Great streets. University of California Transportation Center; 1993.

53. Rapoport A. History and precedent in environmental design. Springer Science & Business Media; 2013.

54. Alexander C. A pattern language: towns, buildings, construction. Oxford university press; 1977.

55. Moudon AV, Lee C. Walking and bicycling: an evaluation of environmental audit instruments. American journal of health promotion. 2003;18(1):21–37.

56. Craig CL, Brownson RC, Cragg SE, Dunn AL. Exploring the effect of the environment on physical activity: A study examining walking to work. Am J Prev Med [Internet]. 2002;23(2, Supplement 1):36–43. Available from: https://www.sciencedirect.com/science/article/pii/S0749379702004725

57. Gehl J. Life between buildings. 2011;

58. Talen E. Pedestrian access as a measure of urban quality. Planning Practice and Research. 2002;17(3):257–78.

59. Mäki N, Sinkko H. Helsingin väestö vuodenvaihteessa 2021/2022 ja väestönmuutokset vuonna 2021 [Internet]. Helsinki; 2022 [cited 2024 May 28]. Available from: https://www.hel.fi/hel2/tietokeskus/julkaisut/pdf/23_01_10_Tilastoja_7_Maki_Sinkko.pdf

60. Norppa M, Hovi H. Kaunis, vihreä ja rauhallinen jalan kaupunginosissa: Asukaskyselyn tulokset [Internet]. Helsinki; 2020 [cited 2024 May 28]. Available from: https://ahjojulkaisu.hel.fi/9ABAF725-8151-C780-95A2-7B7346000000.pdf

61. Bardhan R, Kurisu K, Hanaki K. Does compact urban forms relate to good quality of life in high density cities of India? Case of Kolkata. Cities [Internet]. 2015;48:55–65. Available from: https://www.sciencedirect.com/science/article/pii/S026427511500089X

62. Soini K, Vaarala H, Pouta E. Residents' sense of place and landscape perceptions at the rural–urban interface. Landsc Urban Plan [Internet]. 2012;104(1):124–34. Available from: https://www.sciencedirect.com/science/article/pii/S0169204611002908

63. Assunção G, Patrão B, Castelo-Branco M, Menezes P. An overview of emotion in artificial intelligence. IEEE Transactions on Artificial Intelligence. 2022;3(6):867–86.

64. Bilgram V, Laarmann F. Accelerating innovation with generative AI: AI-augmented digital prototyping and innovation methods. IEEE Engineering Management Review. 2023;

65. Bruce N, Condie J, Henshaw V, Payne SR. Analysing olfactory and auditory sensescapes in English cities: Sensory expectation and urban environmental perception. Ambiances Environnement sensible, architecture et espace urbain. 2015;

66. Thibaud JP. The sensory fabric of urban ambiances. The Senses and Society. 2011;6(2):203–15.

67. Gjerde M. Visual aesthetic perception and judgement of urban streetscapes. In: Paper for Building a Better World: CIB World Congress. Citeseer; 2010. p. 12–22.

**Appendix A – Preliminary analysis**

In our preliminary evaluation to determine the best model for our study, we explored using BLIP, CLIP, and GPT-4.

As CLIP's primary strength lies in image classification (e.g., categorizing an image as 70% park, 20% residential area), it was not suitable for directly scoring visual appeal. We attempted to classify images based on various criteria such as greenery, pedestrian paths, building types, public amenities, among others. Using the classifications provided by CLIP, we created sentences to describe the spaces and then applied sentiment analysis to these sentences. However, this approach was flawed because sentiment analysis models were not effectively measuring the visual appeal of the described space but rather the sentiment of the sentence itself. When we used GPT-4 to interpret the visual appeal of these sentences instead of a sentiment analysis model, the results were not sensible. The sentences generated by CLIP failed to capture many aspects of the images, resulting in a mere report of image segmentation.

In contrast, BLIP proved more effective in generating captions and answering questions. We used a variant called BLIP_VQA from the BLIP model family. By having BLIP answer higher-level questions about the images rather than merely classifying them, we generated more meaningful sentences to analyze in GPT-4. Despite minor grammatical errors, ChatGPT could understand and process these statements effectively. Here's an example of the statement we provided to ChatGPT:

"The roads and pedestrian paths are cracked and worn. The greenery in the area is it is sparse. The types of buildings visible are primarily modern and traditional. The area is predominantly rural. Public amenities such as benches and lighting fixtures are present. Accessibility for different abilities is adequate. Identifiable safety features include crosswalk. The area integrates with adjacent neighborhoods or landmarks yes. There are structures or areas with local historical or cultural significance. The area does not offer a mix of commercial, residential, and recreational spaces. Sustainable features like solar panels or eco-friendly integrations are present. Designated spaces for social interaction are not available."

Our approach using GPT-4 alone was explained in the main text, so we will not discuss it further here.

For our preliminary analysis, we also asked 10 individuals (all non-residents) to rate 10 pictures to compare their ratings with those generated by the AI models. The results showed that the average difference between the ratings of using GPT-4 only was 0.56, whereas using the combination of BLIP and GPT-4 was 0.76. Consequently, we decided to use GPT-4 only for the main study. This decision was based on the finding that GPT-4 provided more consistent and sensible ratings without the need for additional preprocessing or the use of multiple models.

## Appendix B – GPT-4 Prompts

Prompt 1

Imagine you are a human resident of Helsinki (OR human tourist in Helsinki), Finland with (OR without) a typical local perspective on aesthetics and environment. Based on the panoramic image provided, rate the overall visual appeal and functionality of this specific location on a scale of 1 (completely unappealing) to 7 (completely appealing).

Please exclude temporary elements such as weather or passing vehicles. Consider the image as if you're experiencing the environment in person and not just as a viewer of a photograph.

You must not provide any rational or any conversation. I only need one integer number between 1 to 7.

Prompt 2

Imagine you are a human resident of Helsinki (OR human tourist in Helsinki), Finland with (OR without) a typical local perspective on aesthetics and environment. Based on the panoramic image provided, rate the overall visual appeal and functionality of this specific location on a scale of 1 (completely unappealing) to 7 (completely appealing). Focus your assessment on the following criteria:

- Sidewalk Features for Pedestrian Activity: Assess the design and features of the sidewalks. Consider aspects like width, surface condition, pedestrian signage, and accessibility features (e.g., curb cuts, tactile paving) that facilitate comfort and activity.

- Street Design for Traffic and Activity: Evaluate the street layout and design. Focus on street width, lane markings, traffic calming measures (e.g., speed bumps, pedestrian crossings), and the integration of cycle paths or public transit stops, assessing how these features impact traffic flow and pedestrian interaction.

- Tree Canopy and Greenery: Consider the presence of greenery and its contribution to the area's ambiance, irrespective of seasonal changes.

- Physical Indicators of Human Activity: Assess features indicating a space designed for human activity, such as street furniture, public space design, and amenities like water fountains and public art, reflecting potential vibrancy and safety.
- Permanent Lighting: Examine the placement and design of lighting fixtures, disregarding temporary effects of natural lighting due to weather conditions.

Please exclude temporary elements such as weather or passing vehicles. Consider the image as if you're experiencing the environment in person and not just as a viewer of a photograph.

You must not provide any rational or any conversation. I only need one integer number between 1 to 7 per criterion in the format of [##, ##, ##, ##, ##].

Prompt 3

Imagine you are a human resident of Helsinki (OR human tourist in Helsinki), Finland with (OR without) a typical local perspective on aesthetics and environment. Based on the panoramic image provided, rate the overall visual appeal and functionality of this specific location on a scale of 1 (completely unappealing) to 7 (completely appealing). Focus your assessment on the following criteria:

Enduring Physical Features:

- Sidewalk Features for Pedestrian Activity: Assess the design and features of the sidewalks. Consider aspects like width, surface condition, pedestrian signage, and accessibility features (e.g., curb cuts, tactile paving) that facilitate comfort and activity.
- Street Design for Traffic and Activity: Evaluate the street layout and design. Focus on street width, lane markings, traffic calming measures (e.g., speed bumps, pedestrian crossings), and the integration of cycle paths or public transit stops, assessing how these features impact traffic flow and pedestrian interaction.
- Tree Canopy and Greenery: Consider the presence of greenery and its contribution to the area's ambiance, irrespective of seasonal changes.

- Physical Indicators of Human Activity: Assess features indicating a space designed for human activity, such as street furniture, public space design, and amenities like water fountains and public art, reflecting potential vibrancy and safety.
- Permanent Lighting: Examine the placement and design of lighting fixtures, disregarding temporary effects of natural lighting due to weather conditions.

Urban Design Qualities:

- Imageability: Determine the visual distinctiveness and memorability of the environment.
- Legibility: Evaluate how easily one can understand and navigate the spatial layout.
- Enclosure: Consider the sense of spatial definition provided by buildings and natural elements.
- Human Scale: Observe how the proportions of space and elements align with human dimensions for comfort.
- Transparency: Assess the visibility and perceived openness of space, including sightlines and visual connections.
- Linkage: Analyze how different spaces within the image are connected to facilitate movement and interaction.
- Complexity: Reflect on the variety and visual richness of the environment.
- Coherence: Judge the consistency and unity of the urban design elements.

Subjective Reaction: Contemplate your instinctive response to the area's appeal, considering the potential for enjoyment and engagement with the space.

Please exclude temporary elements such as weather or passing vehicles. Provide a balanced assessment without leaning towards an overly positive or negative evaluation. Consider the image as if you're experiencing the environment in person and not just as a viewer of a photograph.

You must not provide any rational or any conversation. I only need one integer number between 1 to 7 per criterion in the format of [##, ##, ##, ##, ##, ##, ##, ##, ##, ##, ##, ##, ##, ##].

**Appendix C - Participant Guidance for Image Rating**

Introduction

Thank you for participating in our urban environmental analysis study. Your insights are valuable in understanding how people perceive urban spaces. This guide will help you focus on the essential aspects of the images you will be rating.

Objective

Our study aims to evaluate the visual appeal of urban spaces using street view imagery. Your task is to rate each image based on the permanent qualities, considering how you would feel if you were physically present in these environments.

What to Focus On

Permanent Features: Pay attention to elements that are constant or long-term in the environment, such as:

- Building architecture and style
- Presence and quality of green spaces (parks, trees, gardens)
- Walkability and pedestrian spaces
- Urban design elements (street layout, benches, lighting)
- General cleanliness and upkeep

What to Ignore

Please disregard temporary or fleeting aspects that do not reflect the inherent qualities of the space, such as:

- Weather Conditions: Sunny, cloudy, rainy, etc.
- Temporary Objects: Passing cars, temporary constructions, movable objects.
- People: Crowds, individuals, or any activities that are not permanent features of the space.

Rating Process

- Imagine yourself in the environment: Consider how you would feel and what your experience would be like if you were there.
- Be consistent: Try to maintain a consistent standard in your ratings throughout the process.

- Trust your instincts: Your first impression is often the most reflective of your true perception of the space.

Conclusion

Your honest and thoughtful ratings are crucial for our study. By focusing on the permanent, inherent qualities of these urban environments, your input will help us create a more accurate and meaningful analysis of urban pleasantness.

**Appendix D - Luminosity**

To determine the luminosity, we apply the formula:

$$L = 0.2126*R + 0.7152*G + 0.0722*B \hspace{2cm} \text{Eq. D.1}$$

where $L$ is luminosity; $R$, $G$, and $B$ are the red, green, blue bands of the image, respectively. This formula is derived from the luminosity function, which reflects how the human eye perceives brightness. This specific calculation is used in converting color images to grayscale, as established in standards like BT.709, which is utilized for HDTV.

A primary reference for this formula is the ITU-R Recommendation BT.709, also known as Rec. 709. This recommendation outlines various parameters for high-definition television, including color representation and luminance coefficients. The coefficients indicate the relative contributions of the red, green, and blue components, respectively, to the perceived brightness. These values are derived from the human visual system's response to these colors.

For a more comprehensive understanding and detailed explanation, refer to:

**Title**: Recommendation ITU-R BT.709-6: Parameter values for the HDTV standards for production and international programme exchange

**Organization**: International Telecommunication Union (ITU)

**Publication Date**: 2015

Following the calculation of luminosity for images, our analysis revealed a weak correlation of -0.16 between luminosity and ratings. Although we initially anticipated a positive correlation, the weak nature of this correlation suggests that it is not significant; therefore, we decided not to pursue further analysis or adjust the ratings based on luminosity.

## Appendix E - Normality Testing of Adjusted Ratings

To test the normality of the ratings, we applied the Shapiro-Wilk test. Although all p-values were significant, the W statistics indicated that the distributions for the first prompts (Model-1 LR and Model-1 NR) were non-normal (Table E.1).

| Ratings | Statistic | P-Value |
|---|---|---|
| Local Residents | 0.99 | 0 |
| Non-Residents | 0.99 | 0 |
| Model-1 LR | 0.70 | 0 |
| Model-2 LR | 0.89 | 0 |
| Model-3 LR | 0.93 | 0 |
| Model-1 NR | 0.69 | 0 |
| Model-2 NR | 0.90 | 0 |
| Model-3 NR | 0.92 | 0 |

Table E.1 - Shapiro-Wilk test results for normality of ratings. (Significance level = 0.05)

**Appendix F - Local Moran's I**

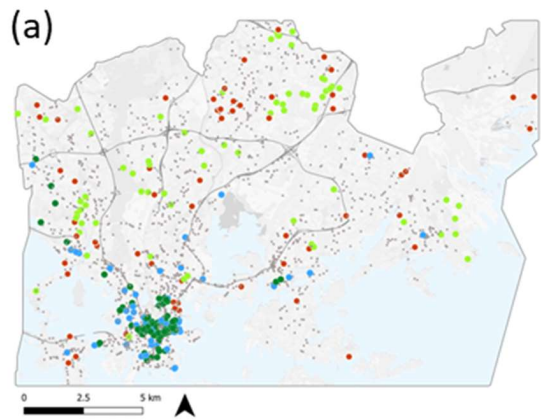
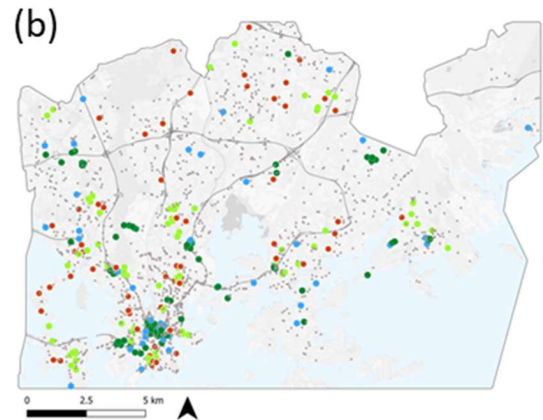
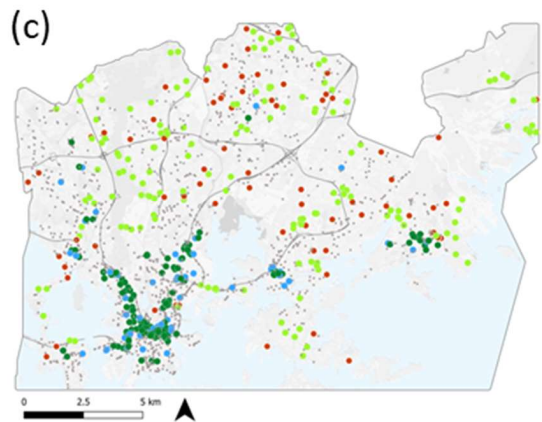
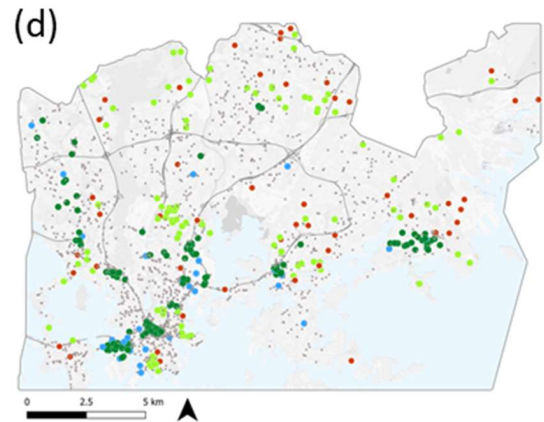
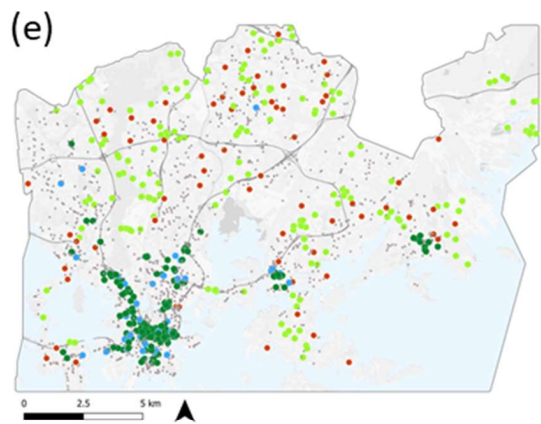
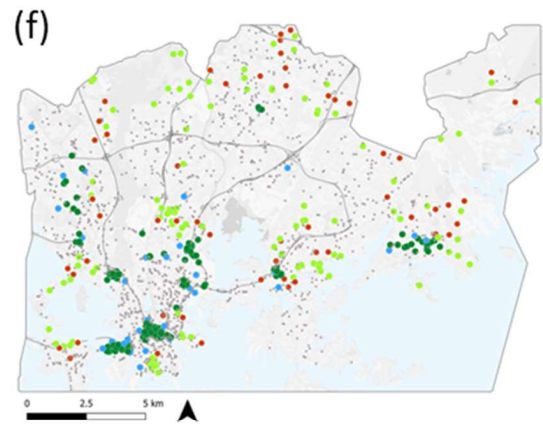

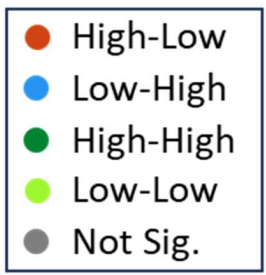

Figure F.1 - Local Moran's I cluster analysis of the differences between GPT-4 model ratings and participant ratings (local residents and non-residents) for various models. Panels (a), (c), and (e) show the results for Model-1, Model-2, and Model-3 respectively, for local residents, while panels (b), (d), and (f) correspond to the same models for non-residents. Red dots represent significant high-high clusters where GPT-4 model ratings are higher than participant ratings and are clustered, blue dots represent significant low-high clusters where participant ratings are higher than GPT-4 model ratings and are clustered, dark green dots indicate significant high-high clusters, light green dots indicate significant high-high clusters, and grey dots indicate areas with no significant clustering.